\documentclass[a4paper,fleqn,usenatbib]{mnras}

\usepackage{newtxtext,newtxmath}

\usepackage[T1]{fontenc}
\usepackage{ae,aecompl}

\usepackage{graphicx}	
\usepackage{amsmath}	
\usepackage{amssymb}	

\title[DSNB from extensive core-collapse simulations]{Diffuse Supernova Neutrino Background from extensive core-collapse simulations of 8--100 ${\rm M}_\odot$ progenitors}

\author[S. Horiuchi et al.]{Shunsaku Horiuchi,$^{1}$\thanks{E-mail: horiuchi@vt.edu}
Kohsuke Sumiyoshi,$^{2}$
Ko Nakamura,$^{3}$
Tobias Fischer,$^{4}$ 
\newauthor
Alexander Summa,$^{5}$ 
Tomoya Takiwaki,$^{6}$
Hans-Thomas Janka,$^{5}$ 
Kei Kotake$^{3,5}$
\\
$^{1}$Center for Neutrino Physics, Department of Physics, Virginia Tech, Blacksburg, VA 24061, USA\\
$^{2}$Division of Liberal Arts, Numazu National College of Technology, Numazu, Shizuoka 410-8501, Japan \\
$^{3}$Department of Applied Physics, Fukuoka University, Fukuoka 814-0180, Japan \\
$^{4}$Institute for Theoretical Physics, University of Wroclaw, pl. M. Borna 9, 50-204 Wroclaw, Poland \\
$^{5}$Max Planck Institut f\"ur Astrophysik, Karl-Schwarzschild-Str. 1, D-85748, Garching, Germany \\
$^{6}$National Astronomical Observatory of Japan, Mitaka, Tokyo 181-8588, Japan \\
}

\date{Accepted XXX. Received YYY; in original form ZZZ}

\pubyear{2016}

\begin{document}
\label{firstpage}
\pagerange{\pageref{firstpage}--\pageref{lastpage}}
\maketitle

\begin{abstract}
We revisit the diffuse supernova neutrino background in light of recent systematic studies of stellar core collapse that reveal the quantitative impacts of the progenitor conditions on the collapse process. In general, the dependence of the core-collapse neutrino emission on the progenitor is not monotonic in progenitor initial mass, but we show that it can, at first order, be characterized by the core compactness. For the first time, we incorporate the detailed variations in the neutrino emission over the entire mass range $8$--$100 M_\odot$, based on (i) a long-term simulation of the core collapse of a $8.8 M_\odot$ ONeMg core progenitor, (ii) over 100 simulations of iron core collapse to neutron stars, and (iii) half a dozen simulations of core collapse to black holes (the ``failed channel''). The fraction of massive stars that undergo the failed channel remains uncertain, but in view of recent simulations which reveal high compactness to be conducive to collapse to black holes, we characterize the failed fraction by considering a threshold compactness above which massive stars collapse to black holes and below which the final remnant is a neutron star. We predict that future detections of the diffuse supernova neutrino background may have the power to reveal this threshold compactness, if its value is relatively small as suggested by interpretations of several recent astronomical observations.
\end{abstract}



\section{Introduction}\label{sec:intro}

At the end of their evolution, massive stars with mass above about $8 M_\odot$ undergo core collapse with the copious emission of neutrinos and anti-neutrinos \citep{Langanke:2002ab,Mezzacappa:2005ju,Kotake:2005zn,Woosley:2006ie,Burrows:2012ew,Janka:2012wk,Foglizzo:2015dma,Janka:2017vlw}. These neutrinos contain rich information that can be exploited to reveal the dynamics and properties of stellar interiors, including the mechanism by which the core collapse reverses to an explosion \citep[for a recent review of supernova neutrinos, see, e.g.,][]{Mirizzi:2015eza}. While a core collapse in our Galaxy will yield revolutionary neutrino data sets \citep[e.g.,][]{Scholberg:2012id}, the occurrence rate is a few per century \citep[e.g.,][]{Diehl:2006cf} and the only remedy is to wait. Future megaton-class neutrino detectors will open the possibility of detecting neutrinos from core collapse in nearby galaxies, with expected returns of a core collapse every few years \citep{Ando:2005ka,Kistler:2008us,Horiuchi:2013bc}. At even larger distances, the combined flux of neutrinos from all past core-collapse supernovae generates the diffuse supernova neutrino background \citep[DSNB; see, e.g.,][for reviews]{Beacom:2010kk,Lunardini:2010ab}. The predicted DSNB flux is tantalizingly close to upper limits currently placed by background-limited searches \citep{Malek:2002ns,Bays:2011si}. Planned upgrades to the Super-Kamiokande detector will dramatically improve signal/background differentiation \citep[][]{Beacom:2003nk} and will open a signal-limited search of the DSNB in the near future. 

A key focus of the DSNB is as a study of the core-collapse neutrino spectrum. Since the DSNB represents the average neutrino emission from all past core collapses, it is sensitive to subpopulations with potentially systematically different neutrino emission properties. A prime example is the contribution from the core collapse to black holes \citep[failed explosions, or sometimes also called un-novae; e.g.,][]{Lunardini:2009ya,Yuksel:2012zy}. The DSNB also serves to test whether an individual event, such as SN1987A---the only core-collapse supernova with neutrino detection to date \citep{Hirata:1987hu,Hirata:1988ad,Bionta:1987qt,Bratton:1988ww}---is typical in its neutrino signal \citep[e.g.,][]{Yuksel:2007mn,Pagliaroli:2008ur,Vissani:2011kx,Vissani:2014doa}. Furthermore, the DSNB has also been discussed as a probe of the cosmic history of the core collapse rate \citep[e.g.,][]{Fukugita:2002qw,Ando:2004sb,Nakazato:2015rya}, as well as a probe of various neutrino physics \citep[e.g.,][]{Ando:2003ie,Fogli:2004gy,Goldberg:2005yw}, which would benefit from a better known neutrino spectrum. 

Predictions of the DSNB have improved steadily over the years \citep{Krauss:1983zn,Dar:1984aj,Totani:1995rg,Totani:1995dw,Malaney:1996ar,Hartmann:1997qe,Kaplinghat:1999xi,Ando:2002zj,Fukugita:2002qw,Strigari:2003ig,Iocco:2004wd,Strigari:2005hu,Lunardini:2005jf,Daigne:2005xi,Yuksel:2005ae,Horiuchi:2008jz,Lunardini:2009ya,Lien:2010yb,Keehn:2010pn,Lunardini:2012ne,Nakazato:2013maa,Mathews:2014qba,Yuksel:2012zy,Nakazato:2015rya,Hidaka:2016zei,Priya:2017bmm}. The neutrino emission predicted from core-collapse simulations can be fitted by a modified Fermi-Dirac spectrum with a spectral shape parameter \citep{Keil:2002in,Tamborra:2012ac}. In DSNB studies it is common to approximate the average neutrino emission from all past core collapses based on such or simpler spectral functions, often fitted to a small selection of core-collapse simulations. Studies of the contribution to the DSNB from core collapse to black holes have likewise utilized spectral functions fitted to a few black hole simulations, and applied across the entire population of stars collapsing to black holes \citep[e.g.,][]{Lunardini:2009ya,Lien:2010yb,Keehn:2010pn,Yuksel:2012zy,Hidaka:2016zei,Priya:2017bmm}. However, in recent years hundreds of core-collapse simulations have been performed in a systematic approach \citep{O'Connor:2010tk,Ugliano:2012kq,Nakamura:2014caa,Pejcha:2014wda,Sukhbold:2015wba,Summa:2015nyk,Ertl:2015rga}. These studies reveal how the neutrino emission quantitatively depends on the progenitor properties. In this paper, we aim to include this dependence in an improved DSNB prediction that better reflects results of ongoing simulation efforts. 

In particular, the so-called compactness of the progenitor has been shown to be a reasonable indicator for the neutrino energetics and many other important quantities in the core collapse. The compactness is a characterization of the stellar core density structure and thus associates with the rate of mass accretion on to the protoneutron star \citep{O'Connor:2010tk}. By strategically choosing its definition (i.e., choosing the appropriate mass $M$ in equation \ref{eq:compactness}), the compactness can capture the mass accretion during the critical epoch of shock revival, and becomes a reasonable predictor of the core-collapse outcome. Indeed, despite being a simplistic one-parameter attempt to describe an inherently multi-dimensional and time evolving phenomenon, a single compactness parameter can predict the outcome of core collapse in more than 80\% of massive stars \citep{Ugliano:2012kq,Pejcha:2014wda,Sukhbold:2015wba}; adopting two parameters yields significantly better accuracy, since both the accretion ram pressure working to halt the explosion and the neutrino heating working to induce the explosion can be captured \citep{Ertl:2015rga}. Furthermore, since mass accretion is the power source of neutrino emission, the compactness can also capture the dependence of the neutrino emission on the progenitor \citep[e.g.,][]{OConnor:2012bsj,Nakamura:2014caa}. 

In this paper, we determine the DSNB by incorporating the impact of the progenitor structure on the neutrino emission by working with the concept of progenitor compactness. To this end, we utilize a suite of 101 axis-symmetric core-collapse simulations of \cite{Nakamura:2014caa} that cover a wide range of progenitor mass. To this we add simulations of core collapse to black holes from the literature \citep{sum07,sum08,Fischer:2008rh,Hempel:2011mk,Fischer:2011cy,Nakazato:2012qf,Huedepohl:2014} as well as run our own black hole forming simulations. Finally, we include a long-term simulation of the core collapse of an ONeMg core \citep{Huedepohl:2009wh}. These cover the range of progenitors necessary for DSNB predictions. We use the compactness of the progenitor to characterize the neutrino emission, and make new predictions for the DSNB by using the distribution of compactness in stellar populations. We determine the expected event rates at current and  future neutrino detectors, and show that the DSNB can reveal the critical compactness that separates black hole forming and neutron star forming collapses, if the transition occurs at relatively small compactness. 

The paper is organized as follows. In Section \ref{sec:sim}, we discuss our core-collapse simulations, for both failed explosions leaving behind black holes (Section \ref{sec:sim:1d}) and core collapse leaving behind neutron stars (Section \ref{sec:sim:2d}). In Section \ref{sec:dsnb}, we discuss our DSNB prediction, starting with our characterization of the neutrino emission in terms of compactness (Section \ref{sec:signal}), followed by our flux (Sections \ref{sec:average} and \ref{sec:effects}) and event (Section \ref{sec:dsnb:signal}) predictions, and discussion of uncertainty arising from the nuclear equation of state (EOS, Section \ref{sec:uncertainties:eos}). We conclude in Section \ref{sec:discussion}. Throughout, we adopt a $\Lambda$ cold dark matter cosmology with $\Omega_m=0.3$, $\Omega_\Lambda=0.7$, and $H_0=70 \, {\rm km \, s^{-1} \, Mpc^{-1}}$ \cite[the `737' cosmology;][]{Rao:2005ab}. 

\section{Simulation setup}\label{sec:sim}

We review the setup of our core-collapse simulations for both core collapse forming black holes and neutron stars. In order to calculate the DSNB, we are primarily interested in obtaining spectral parameters for the time-integrated neutrino emission: the total energetics, mean energies, and spectral shape parameter. Since the details of the weak-interaction physics, nuclear physics, hydrodynamic treatment, and gravity all affect the neutrino spectrum, a detailed numerical treatment is necessary. However, systematic simulations of stellar core collapse spanning the necessary range in progenitors in full three-dimensional (3D) geometry and detailed microphysics are currently not feasible. We thus employ a suite of spherically symmetric one-dimensional (1D) general relativistic simulations to explore collapse to black holes, and a suite of axis-symmetric two-dimensional (2D) simulations to explore the shock revival leading to the possibility of explosion leaving behind neutron stars. We emphasize that it is important to keep in mind that both simulation suites will need to be updated in the future by full 3D simulations. Our foremost aim is to demonstrate qualitative conclusions, bearing in mind predictions will be quantitatively refined as more simulations become available. We discuss the some specific caveats in Sections \ref{sec:sim:compare} and \ref{sec:discussion}. 

\subsection{Spherically symmetric simulations}\label{sec:sim:1d}

Numerical simulations of failed explosions---collapse of massive stars to black holes---are performed by a general relativistic neutrino-radiation hydrodynamics code \citep{yam97,sum05}. The code solves the Boltzmann equation for neutrinos by the S$_n$ (multi-angle) method coupled with the equations of Lagrangian hydrodynamics in general relativity. The multi-energy treatment of the code enables us to derive detailed information of neutrino energy spectra for all neutrino species. A fully implicit method for time evolution is advantageous to follow the long-term evolution ($\sim 1$ sec) of compact objects born after the core bounce. 	The code has been applied to the study of core-collapse supernovae \citep{sum05} and black hole formation in massive stars \citep{sum06,sum07,sum08}.  

We implement nuclear physics in the same way as in \cite{sum05,sum07}. The basic set of neutrino reactions rates \citep{bru85} for emission, absorption, and scattering processes as well as pair-processes and the nucleon--nucleon bremsstrahlung are taken into account. The code solves the neutrino distributions of four species ($\nu_e$, $\bar{\nu}_e$, $\nu_\mu$ and $\bar{\nu}_{\mu}$; the latter two of which also represent $\nu_{\tau}$ and $\bar{\nu}_{\tau}$). The equation of state (EOS) by \cite{Lattimer:1991nc} with the incompressibility of 220 MeV is adopted in the current study. 
It is worth mentioning that the uncertainties in nuclear physics needs to be further scrutinized, including more sophisticated neutrino reaction rates \citep[e.g.,][]{len12a,MartinezPinedo:2012rb,Martinez-Pinedo:2013jna,Horowitz:2016gul} and modern set of EOSs \citep{2016arXiv161003361O}. We will discuss the influence of EOS and our EOS choice below in \S \ref{sec:uncertainties:eos}. 

In order to reveal the systematics of neutrino signals from a variety of massive stars, we choose the compactness parameter,
\begin{equation} \label{eq:compactness}
\xi_M = \left. \frac{M/M_\odot}{R(M_{\rm bary}=M)/1000\,{\rm km}} \right\vert_{ t },
\end{equation}
where $R(M_{\rm bary}=M)$ is the radial coordinate that encloses a baryonic mass $M$ at epoch $t$, to characterize the density profiles of the progenitors. In particular, the compactness defined by a large mass of $M \approx 2.5 M_\odot$ is useful to estimate the time period from the core bounce till the black hole formation and the duration of neutrino emission \citep{O'Connor:2010tk}. Since the iron core mass is large and the accretion rate of the outer layer is high for a large compactness, a newly born protoneutron star at the center rapidly becomes massive and leads to dynamical collapse to a black hole when it reaches the critical neutron star mass. 

We thus set up initial conditions taken from the central core of massive stars selected to cover a range of compactness. We follow the time evolution of gravitational collapse, core bounce with a launch of shock wave, and its stall due to the accretion of matter. As we will see, the compactness controls the accretion rate through the free-fall time, the time duration till the black hole formation and, therefore, the energetics of neutrino emission. We cover a range of the compactness by choosing the following progenitor models: 40 M$_{\odot}$ of \cite{Woosley:1995ip} (WW95) with solar metallicity, and models of 33M$_{\odot}$ and 35M$_{\odot}$ of \cite{Woosley:2002zz} (WHW02) with zero metallicity. These progenitors have a compactness of $ \xi_{2.5} = 0.55$, 0.39 and 0.52, respectively. Here and elsewhere, we evaluate the compactness of the progenitor \citep{Sukhbold:2013yca} instead of the original definition at $t$ of core bounce \citep{O'Connor:2010tk}. 

Radial grids of the mass coordinate are arranged in a non-uniform way to cover the central objects, the shock wave and the accreting material.  
The numbers of grids for radial mass coordinate, neutrino angle and energy are 255, 6 and 14, respectively. In order to describe the long-term evolution of accretion we utilize the rezoning of grids during the calculations.  For cases of low accretion rates, we utilize a fine mesh with 511 grids for the radial mass coordinate. Since the case of $40 M_{\odot}$ is available from previous studies \citep{Nakazato:2008vj,Nakazato:2012qf}, we newly performed the two cases of $33M_{\odot}$ and $35M_{\odot}$ for the current study. Note that our numerical simulations can provide detailed information of neutrino spectra by solving the Boltzmann equations, being different from the survey by \cite{O'Connor:2010tk}. The library of neutrino signal for various progenitors can be found in \cite{Nakazato:2012qf}.  

We evaluate the neutrino's spectral parameters utilizing the neutrino distribution function for neutrino energy, $\varepsilon_{\nu}$, $f_{\nu}(\mu_{\nu}, \varepsilon_{\nu}, t)$ at the surface of the core, obtained from the Boltzmann equation, where $\mu_{\nu}=\cos \theta_{\nu}$ is cosine of the neutrino propagation angle with respect to the radial coordinate. The neutrino luminosity, mean energy, and shape parameter are defined by,
\begin{eqnarray} \label{eq:1D_Neutrino_Averages}
L_{\nu}(t) &=& 4\pi R_s^2 c F_{\nu}(t) ~, \\
E_\nu(t) &=& \langle \epsilon_{\nu}(t) \rangle ~ = ~ \frac{\epsilon_{\nu}(t)} {n_{\nu}(t)}~, \\
\alpha(t)  &=& \frac{2\langle\epsilon_{\nu}\rangle^2 - \langle \epsilon_{\nu}^2 \rangle}
                    {\langle \epsilon_{\nu}^2 \rangle - \langle \epsilon_{\nu} \rangle^2} ~,
\end{eqnarray}
where $R_s$ is taken as the iron core radius (for our chosen progenitors, in the range 0.6--$1 \times 10^9$ cm) and the mean-squared energy is,
\begin{eqnarray} \label{eq:1D_Neutrino_Energies}
\langle \epsilon_{\nu}^2 (t) \rangle &=& \frac{\epsilon_{\nu}^2(t)}  {n_{\nu}(t)} ~,
\end{eqnarray}
and the average energies and flux are evaluated using energy and angle moments of neutrino distribution function by,
\begin{eqnarray} \label{eq:1D_Neutrino_Moments} 
         n_{\nu}(t) &=& \int \frac{d^3p_{\nu}}{(2\pi \hbar c)^3} f_{\nu} ~, \\
\epsilon_{\nu}(t)   &=& \int \frac{d^3p_{\nu}}{(2\pi \hbar c)^3} f_{\nu} \varepsilon_{\nu} ~, \\
\epsilon_{\nu}^2(t) &=& \int \frac{d^3p_{\nu}}{(2\pi \hbar c)^3} f_{\nu} \varepsilon_{\nu}^2 ~, \\
F_{\nu}(t) &=& \int \frac{d^3p_{\nu}}{(2\pi \hbar c)^3} f_{\nu} \mu_{\nu} \varepsilon_{\nu} ~, 
\end{eqnarray}
through the integration of neutrino momentum space, i.e., neutrino energy $\varepsilon_{\nu}$ and angle $\mu_{\nu}$. 

\begin{figure}
\includegraphics[width=125mm,bb=0 50 800 580]{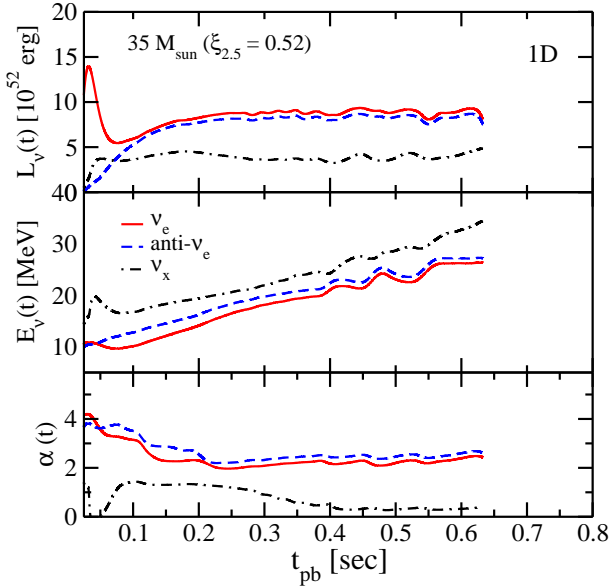}
\caption{Time evolution of neutrino spectral parameters for the core collapse of the $35M_{\odot}$ progenitor leading to black hole formation at 630 msec post bounce. The neutrino luminosity (top panel), mean energy of neutrinos (middle panel), and shape parameter $\alpha$ (bottom panel) are shown for $\nu_e$ (red solid), $\bar{\nu}_e$ (blue dashed), and $\nu_\mu$ (black dot--dashed); the $\bar{\nu}_{\mu}$ are not shown for clarity but are quantitatively very similar to $\nu_\mu$. All quantities are shown as functions of time after the core bounce. }
\label{fig:BHneutrino}
\end{figure}

We show in Figure \ref{fig:BHneutrino} the three spectral parameters (luminosity, mean energy, and shape parameter) of neutrinos emitted from the collapse of the $35M_{\odot}$ star as an example. The duration of neutrino burst is short, $\sim$ 630 msec for this case, from core bounce until the termination due to black hole formation. The rapid increase of the mean energies of all neutrino species during the short burst is the hallmark signature of the evolution towards black hole formation. As the protoneutron star grows massive due to mass accretion, it becomes compact with increasing density and temperature. Accordingly, the energies of neutrinos rapidly increase from the moment of core bounce until the formation of the back hole. 

\begin{figure}
\includegraphics[width=125mm,bb=0 50 800 580]{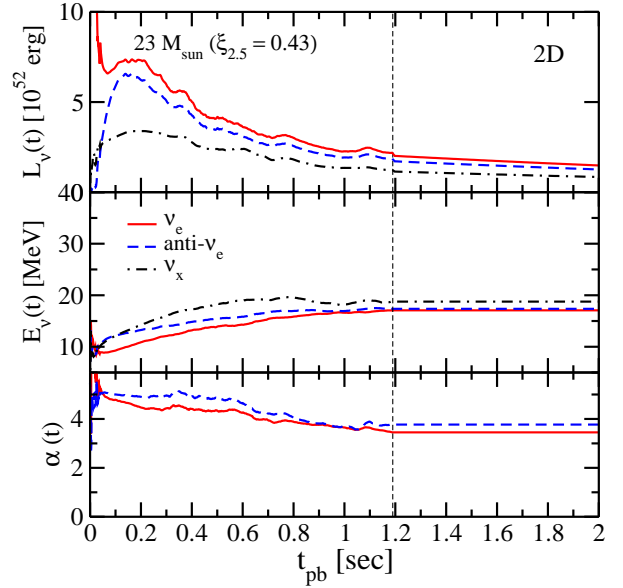}
\caption{The same as Figure \ref{fig:BHneutrino}, but for 2D simulation. The $23 M_\odot$ progenitor is chosen for comparison, since it has the highest $\xi_{2.5}$ among the 101 solar metallicity progenitors of the WHW02 suite used in this study. The vertical dashed line shows the transition from numerical hydrodynamic to analytic extrapolation regimes.}
\label{fig:NSneutrino}
\end{figure}

The initial behaviour during core bounce is similar to the ordinary case of collapse to neutron stars, i.e., neutrinos showing the usual peaks due to the neutronization burst (in $\nu_e$) and the passage of shock wave. As in the neutron star case, the $\nu_e$ and $\bar{\nu}_e$ luminosities originate from the energy release by neutrinos through electron (positron) absorptions in the accreting matter and show variations according to the accretion rate. A peak in the energy of $\nu_\mu$ is observed around the timing of the neutronization burst, which is due to the passage of the shock wave through neutrinospheres. High-energy neutrinos are created at high temperature due to the shock passage right after core bounce. Those neutrinos outwardly propagate from the neutrino {\it thermal} sphere and remain without degrading energy until they are emitted from neutrino {\it scattering} sphere. This brief hardening of spectra leads to a temporary drop of the shape parameter. This phenomenon has also been seen in previous studies \citep{Liebendoerfer:2003es,Buras:2005rp,Lentz:2012xc} and is not seen with energy changing reactions.

\subsection{Axis-symmetric simulations}\label{sec:sim:2d}

We adopt two sets of 2D axis--symmetric core-collapse models. In these 2D models, the same EOS (LS EOS with $K=220$ MeV) was adopted as in our 1D models, while self-gravity and neutrino transport were solved in different ways. The first set of simulations we adopt is from \cite{Nakamura:2014caa}. In these models, self-gravity was computed with a Newtonian monopole approximation, and neutrino transport for electron and anti--electron neutrinos ($\nu_{\rm e}$ and $\bar{\nu}_{\rm e}$) are performed with an energy-dependent treatment of neutrino transport based on the isotropic diffusion source approximation \cite[IDSA;][]{Liebendoerfer:2007dz} with a ray-by-ray approach. This approximation has a high computational efficiency in parallelization, which allows us to explore systematic features of neutrino emission for a large number of supernova models. Regarding heavy--lepton neutrinos ($\nu_x = \nu_\mu,\, \nu_\tau,\, \bar{\nu}_\mu,\, \bar{\nu}_\tau$), a leakage scheme was employed to include cooling processes. Since the leakage scheme does not enable us to obtain spectral information, we assume that the average energy of $\nu_x$ is given by the temperature of matter at the corresponding average neutrinosphere.

In \cite{Nakamura:2014caa}, 378 non-rotating progenitor stars from WHW02 covering zero-age main sequence (ZAMS) mass from 10.8 $M_\odot$ to 75 $M_\odot$ with metallicity from zero to solar value were investigated. From these, we choose 101 supernova models with solar metallicity for the current study. This is because lower metallicity supernovae are dominant in distant galaxies where the neutrinos would suffer from energy redshift and thus contribute little to the detectable DSNB signal. The chosen 101 models cover a wide range of compactness ($\xi_{2.5}$ from 0.0033 for the 10.8 $M_\odot$ model to 0.434 for the 23.0 $M_\odot$ model).

The neutrino luminosity, mean energy and the shape parameter are defined in the same manner as in our 1D models (except $\nu_x$ mean energy and shape parameter as described above). Since the 2D simulations in \cite{Nakamura:2014caa} were terminated at $\sim 1$ s after bounce, the neutrino emissions during the cooling phase of the protoneutron star were not fully solved. More than half of the total neutrino energy is expected to be emitted after the end of the simulations. Thus, we extrapolate the neutrino properties based on some assumptions: 
(1) the core radius contracts according to $R(t) = R_{\rm f} + (R_{\rm i}- R_{\rm f}) {\rm e}^{-t/t_0}$ \citep{Arcones07}, where the final core radius $R_{\rm f}$ is set to be 15 km,  and the initial radius $R_{\rm i}$ and the contraction time-scale $t_0$ are found by fitting this function to the time evolution of the core radius during the simulations; 
(2) the gravitational energy released per unit time by the core contraction is converted to the loss rate of neutrino energy (total neutrino luminosity) with a conversion efficiency $\beta$ which is of the order of unity and determined at the final time of each simulation; and
(3) the average neutrino energies and relative luminosity ratios among neutrino flavours are fixed to the final value of each simulation. In this extrapolation process we ignore the contribution of accreting matter to the neutrino emission, which is a good assumption for the late phase of protoneutron star evolution. Long-term core-collapse simulations taking into account energy-dependent neutrino transport and core evolution will improve our supernova neutrino models in the future.

The second set of 2D simulations is from \cite{Summa:2015nyk}. In these models, self-gravity is computed using general relativistic monopole corrections as described in \cite{Marek:2005if}, and neutrino transport is solved with a ray-by-ray approximation along radial rays using a variable Eddington factor method \citep[e.g.,][]{Buras:2005rp} for all neutrinos. A total of 18 progenitor models in the ZAMS mass range $11.2$--$28 M_\odot$ are selected from WHW02 and \cite{Woosley:2007as}, spanning compactness $\xi_{2.5}$ from 0.005 (for the 11.2 $M_\odot$ model) to 0.33 (for the 25.0 $M_\odot$ model). Since the number of progenitor is smaller, we mainly adopt the simulation suite of \cite{Nakamura:2014caa} for this paper; however, the \cite{Summa:2015nyk} models serve as an excellent comparison to see potential systematic differences due to numerical treatments. 

We compute the neutrino luminosity, mean energy and the shape parameter as defined in the same manner as in our previous models with one modification. Since the \cite{Summa:2015nyk} models were terminated typically at $\sim 0.5$ s after bounce, there is less post-accretion phase data to perform a reliable fit to the late-time radius evolution following the prescription of \cite{Arcones07}. Furthermore, in some cases the protoneutron mass is still visibly increasing at the final simulation time step. We therefore adopt the following assumptions to compute the late-time emission, noting that the emphasis is to obtain time-integrated quantities: (1) the core radius contracts from the radius at the final time step to a final radius of 15 km; (2) the mass grows according to $M(r) = M_0 + M_1(1- e^{-t/\tau_M})$, where $M_0$, $M_1$, and $\tau_M$ are found by fitting this function to the time evolution of the protoneutron mass; (3) the gravitational binding energy released after the final simulation step is equipartitioned between all neutrino flavours; and (4) the average neutrino energies and pinching factors are fixed to the final value of each simulation. 

\subsection{Similarities and differences between 1D and 2D}\label{sec:sim:compare}

We have utilized 1D and 2D core-collapse simulations to model core collapse to black holes and neutron stars, respectively. Ideally, we want any differences between the simulation suites to reflect only the different outcomes of core collapse. However, the different implementations of microphysics and numerical treatments can also affect the neutrino emission. While the simulation suites share many core microphysics implementations, the list of what are considered necessary interactions and their implementations are topics of ongoing research. Here, we compare and contrast the results of our 1D and 2D simulations, but this caveat should be kept in mind. Future systematic studies, informed by state-of-the-art focused simulations, will continuously improve predictions. 

Sample neutrino emission properties are shown in Figures \ref{fig:BHneutrino} and \ref{fig:NSneutrino} for collapse to black holes and neutron stars, respectively. Figure \ref{fig:BHneutrino} shows a $35 M_\odot$ progenitor with compactness $\xi_{2.5} \approx 0.52$. Since there is no progenitor in our 2D suites with such a high compactness, we compare with the the largest available: the $23 M_\odot$ progenitor with $\xi_{2.5} = 0.43$. The initial bounce phase is similar in both 1D and 2D: a strong neutronization burst signal (in $\nu_e$), followed by the rise in $\bar{\nu}_e$ and $\nu_x$ emissions; and the expected hierarchy in neutrino luminosities ($\nu_x < \bar{\nu}_e < \nu_e$) and mean energies ($\nu_e < \bar{\nu}_e < \nu_x$). However, the failed explosion necessarily experiences high post-bounce mass accretion, which drives the protoneutron star above its mass limit (at $\sim 630$ msec post bounce for the $35 M_\odot$ progenitor). By contrast, the shock revival experienced in the 2D simulation leads to dramatically reduced mass accretion. The difference in mass accretion is responsible for significantly larger neutrino luminosities and energies in the failed explosion. Progenitors with lower compactness typically show lower luminosities with shorter accretion phase durations, and slightly lower mean energies \cite[see, e.g.,][]{OConnor:2012bsj}. Our simulations demonstrate the same dependences. In Figure \ref{fig:nu}, we show how they translate to dependences for the time-integrated neutrino emission parameters.

Differences can also be seen in the spectral shape parameter $\alpha$. Systematically smaller values are obtained for failed explosions for $\nu_e$ and $\bar{\nu}_e$ (we cannot compare the shape parameter of $\nu_x$ due to a lack of detailed transport for $\nu_x$ in Nakamura's 2D simulations). Comparing flavours, the $\nu_x$ shows a significantly smaller shape parameter than $\nu_e$ and $\bar{\nu}_e$. To see whether these differences are real or numerical, we can compare the results of our 2D simulation with that in Figure 10 of \cite{Mirizzi:2015eza}, where the core collapse of a $27 M_\odot$ progenitor was simulated using LS220 EOS. They find the shape parameter decreases, from around 4--5 to $\sim 3$ for $\nu_e$ and $\bar{\nu}_e$, and from around $\sim 3$ to $\sim 2$ for $\nu_x$, during the first second post bounce. This behaviour confirms that indeed the $\nu_x$ should have lower shape parameters than $\nu_e$ or $\bar{\nu}_e$. This is primarily driven by the $\nu_x$ emission arising from deeper in the protoneutron star with a larger neutrinosphere width, which results in the contribution of higher temperature emissions and hence a smaller $\alpha$. Thus, we conclude that it is reasonable that the shape parameter shows an even larger suppression in failed explosions where the larger accretion rates heat the protoneutron star and neutrinospheres. Additionally, as a more detailed comparison we also ran a simulation of the $27 M_\odot$ progenitor with LS220 EOS using a three-flavour IDSA transport scheme (Takiwaki et al., in prep). We found that $\alpha$ for $\nu_e$ and $\bar{\nu}_e$ starts around 4--5 and drops to 3--4 by $\sim 1$ sec post bounce, while the $\nu_x$ falls from $\sim 2$ to $\sim 1$, similar in behaviour to the results of \cite{Mirizzi:2015eza}. Since the high energy component of the neutrino distribution is essential to determine the shape parameter, the inclusion of improved neutrino reactions such as inelastic scatterings on nucleons and nuclei, which contribute to down-scattering of high energy neutrinos, is important for the detailed prediction of the neutrino spectra. 

\section{DSNB prediction and detection}\label{sec:dsnb}

\subsection{Characterizing the neutrino emission}\label{sec:signal}

For each 2D core-collapse simulation, we estimate the total energy liberated in the form of neutrinos and the flux-weighted mean neutrino energy as, 
\begin{eqnarray}\label{eq:parameters}
E^{\rm tot}_\nu &=& \int L_\nu(t) dt~,  \\
\langle E_\nu \rangle &=& \frac{\int E_\nu(t) \dot{N}_\nu(t)dt}{\int \dot{N}_\nu(t)dt}~, \label{eq:weightedEave}
\end{eqnarray}
where $\dot{N}_\nu = L_\nu / E_\nu$. The time integrals are performed until $100$ sec post bounce. We obtain the spectral shape parameter $\langle \alpha \rangle$ by fitting the time-summed neutrino spectrum to the pinched Fermi-Dirac functional form \citep{Keil:2002in},
\begin{eqnarray}\label{eq:pinchedFD}
f(E)= \frac{(1+  \langle \alpha \rangle )^{(1+  \langle \alpha \rangle )}}{\Gamma(1+  \langle \alpha \rangle )} \frac{E^{\rm tot}_\nu E^{ \langle \alpha \rangle }}{\langle E_\nu \rangle^{2+  \langle \alpha \rangle }} {\rm exp}\left[ {-(1+  \langle \alpha \rangle ) \frac{E}{ \langle E_\nu \rangle }} \right]~,
\end{eqnarray}
where $\langle E_\nu \rangle$ is fixed to the flux-weighted average neutrino energy, equation (\ref{eq:weightedEave}). This yields a better spectral fit than using a flux-weighted shape parameter analogous to the mean energy of equation (\ref{eq:weightedEave}). 

\begin{figure}
\includegraphics[width=125mm,bb=0 50 800 580]{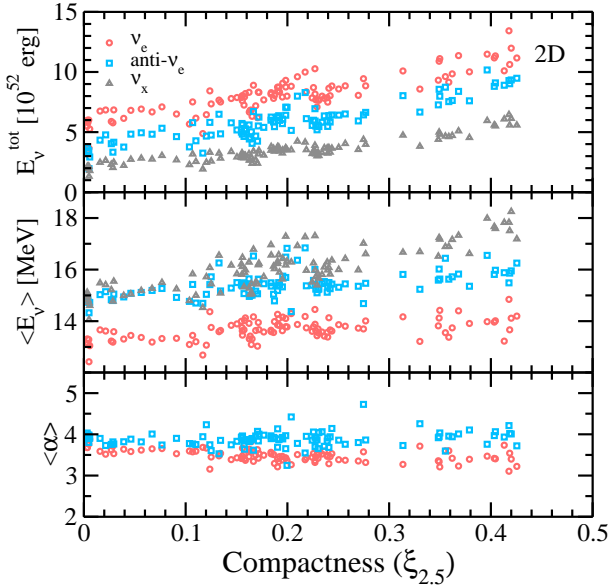}
\caption{Neutrino emission parameters for 2D simulations: total energetics (upper panel), mean energy (middle panel), and shape parameter (lower panel) shown for $\nu_e$ (red circles), $\bar{\nu}_e$ (blue squares), and $\nu_x$ (black diamonds). Each point is a core-collapse simulation, plotted at the value of the progenitor compactness $\xi_{2.5}$.}
\label{fig:nu}
\end{figure}

Figure \ref{fig:nu} shows the spectral parameters $(E^{\rm tot}, \langle E_\nu \rangle, \langle \alpha \rangle)$ separately for $\nu_e$, $\bar{\nu}_e$, and $\nu_x$, plotted as functions of the compactness $\xi_{2.5}$ of the progenitor. The total neutrino energetics show a clear increase with the progenitor compactness, consistent with previous studies: high compactness leads to higher mass accretion rate, which leads to larger gravitational energy liberation and hence higher neutrino energetics \citep{Nakamura:2014caa}. We also see a clear hierarchy in the total energetics in neutrino flavour, $\nu_x < \bar{\nu}_e < \nu_e $. The $\nu_e$ is largest due to the additional contribution from the deleptonization of the progenitor core. For mean energy, a mild increase with compactness is evident. We also see the usual hierarchy with $\nu_e$ being the lowest and $\nu_x$ being the highest. The shape parameter $\langle \alpha \rangle$ shows little variation between different compactness or flavour, consistently falling between 3 and 4. As noted in section \ref{sec:sim:2d}, we do not determine $\langle \alpha \rangle$ for $\nu_x$ due to limitations in our simulation setup. However, as discussed in section \ref{sec:sim:compare}, we expect it to be smaller than the values for $\nu_e$ or $\bar{\nu}_e$, and we will explore values between 1.0 and 4.0 in this paper. 

\begin{figure}
\includegraphics[width=125mm,bb=0 50 800 580]{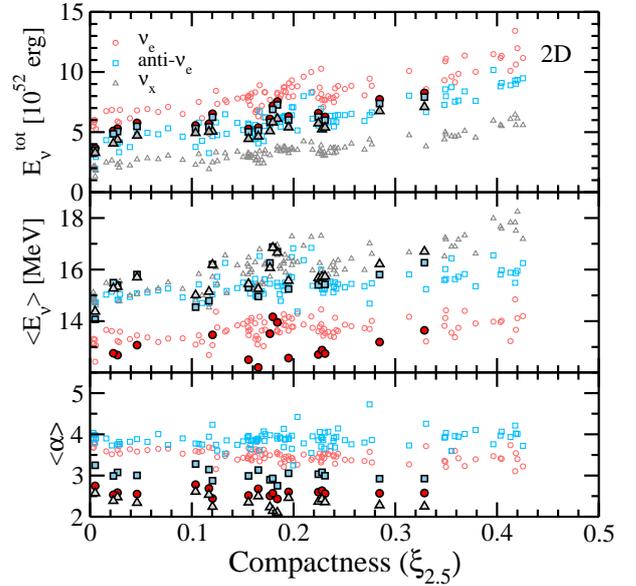}
\caption{The same as Figure \ref{fig:nu} but showing in bold filled symbols the time-integrated spectral parameters for the simulations of Summa et al.~(2016).}
\label{fig:nu2}
\end{figure}

In Figure \ref{fig:nu2} we plot on the same data the time-integrated spectral parameters for the 18 2D core-collapse simulations of \cite{Summa:2015nyk}. The same qualitative trends are observed. For example, the hierarchy in the mean energy and alpha with neutrino species is observed. Most importantly, the same trends with compactness are observed: rise in the total neutrino energetics, mild increase in neutrino average energies, and weak dependence for the shape parameter. The flavour hierarchy of the total energetics is more modest in the \cite{Summa:2015nyk} models by construction: since the \cite{Summa:2015nyk} simulations terminate earlier and we assume the gravitational binding energy liberated in the late-phase is equipartitioned in neutrino species. One notable difference is seen in the systematically lower shape parameters. This is not surprising since as discussed in Section \ref{sec:sim:compare}, the shape parameter is particularly sensitive to the details of the microphysics included.

We next repeat the exercise for collapse to black holes, shown as solid symbols in Figure \ref{fig:nuBH}. For these, the time integral is performed until the moment of black hole formation. For consistency, we collect simulations from the literature that adopt an LS EOS with $K=220$ MeV. \cite{Hempel:2011mk} simulated the collapse of the solar metallicity $40 M_\odot$ star of WW95 updated in \cite{Heger:2000su}, whose $\xi_{2.5} \approx 0.59$. \cite{Nakazato:2012qf} followed black hole formation in their low metal $30 M_\odot$ progenitor, whose compactness is $\xi_{2.5} \approx 0.74$ (private communication). \cite{Huedepohl:2014} simulated the collapse of the solar metal $25 M_\odot$ and $40 M_\odot$ progenitors of WHW02, with compactness $\xi_{2.5} \approx 0.31$ and $\xi_{2.5} \approx 026$, respectively. 

To this list we add simulations of the solar metallicity $40 M_\odot$ star of WW95 with $\xi_{2.5} \approx 0.55$ \cite[note the different compactness from the version updated in][]{Heger:2000su}, as well as the $33 M_\odot$ and $35 M_\odot$ zero-metallicity progenitors of WHW02 with $\xi_{2.5} \approx 0.39$ and $0.52$, respectively. Due to the limited number of black hole forming collapse simulations, we fit the neutrino spectral parameters by a linear function of $\xi_{2.5}$, separately for $\nu_e$, $\bar{\nu}_e$, and $\nu_x$. These are shown by the solid lines in Figure \ref{fig:nuBH}. We find the neutrino energetics tend to decrease with compactness, which can be understood by the fact that larger compactness progenitors have shorter durations for black hole formation; it is well approximated as $\propto \xi^{-3/2}$ \cite[e.g.,][]{O'Connor:2010tk}. Although the shorter duration is partially compensated by higher luminosities, the duration dominates the overall effect. The neutrino mean energies show a flat (for $\nu_e$ and $\bar{\nu}_e$) or rise (for $\nu_x$) with compactness. The rise in $\nu_x$ mean energy is the hallmark signature of black hole formation: the rapid mass accretion increases the protoneutron star density and temperature, which is reflected in $\nu_x$; the other flavours by comparison have larger neutrinosphere radii and are less affected. For example, the $\nu_e$, $\bar{\nu}_e$, and $\nu_x$ neutrinosphere radii are 60, 58, and 30 km for the neutrino energy of 34 MeV at 400 msec after bounce in the case of $35 M_{\odot}$.  The protoneutron star radii shrink to $\sim$30 km ($\rho> 10^{10}$ g/cm$^3$) by this time. The shape parameter shows a weak tendency to increase with compactness. This can be understood by the fact that the duration of the neutrino emission decreases with compactness. Since the shape parameter initially starts with high values and plateaus at low values in the late phase, a shorter emission duration results in a larger value of the time-integrated shape parameter.  

\subsection{IMF-weighted average neutrino spectrum}\label{sec:average}

The average neutrino spectrum per core collapse, $dN/dE$, can be derived by evaluating the contribution of a given progenitor by the initial mass function (IMF). Since the IMF falls steeply with progenitor mass, contributions from lower mass progenitors become important. We use the ZAMS mass bins used in WHW02, which runs from $10.8 M_\odot$ to $75 M_\odot$. Progenitors with initial masses below this range evolve and collapse as ONeMg cores \citep{Jones:2013wda}, and are an important source of the DSNB \citep[e.g.,][]{Mathews:2014qba}. We include this population based on the long-term core-collapse simulation of \cite{Huedepohl:2009wh} who used the $8.8 M_\odot$ progenitor of \cite{Nomoto:1984,Nomoto:1987}. The time-integrated neutrino spectral parameters (not shown in Figure \ref{fig:nu}) are $(E_\nu^{\rm tot}, \langle E_\nu \rangle, \langle \alpha \rangle) = (2.07,8.05,2.53)$, $(2.08,9.13,1.89)$, and $(2.32,9.83,1.43)$ for $\nu_e$, $\bar{\nu}_e$, and $\nu_x$, respectively, in the same units as shown in Figure \ref{fig:nu}. For progenitors above $75 M_\odot$, we extend the $75 M_\odot$ mass bin to $100 M_\odot$. 

\begin{figure}
\includegraphics[width=125mm,bb=0 50 800 580]{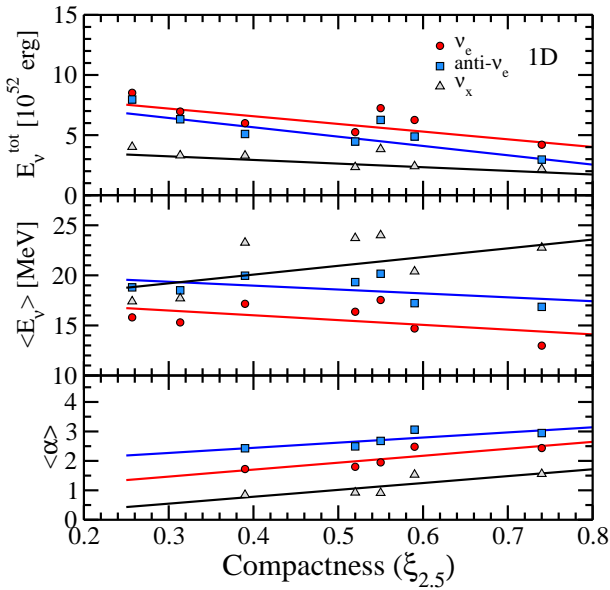}
\caption{Same as Figure \ref{fig:nu} but for failed explosions. Note the different ranges in both horizontal and vertical axes. From small to large compactness, points represent the solar metal $40 M_\odot$ of WHW02, solar metal $25 M_\odot$ of WHW02, zero metal $33 M_\odot$ of WHW02, zero metal $35 M_\odot$ of WHW02, solar metal $40 M_\odot$ of WW95, solar metal $40 M_\odot$ of WW95 updated in Heger et al.~(2001), and low metal $30 M_\odot$ of Nakazato et al.~(2013). Spectral shape parameter are shown only when relevant information was available. Straight lines are linear fits through the spectral parameters.}
\label{fig:nuBH}
\end{figure}

The average neutrino spectrum is then obtained as,
\begin{equation} \label{eq:average}
\frac{dN}{dE} = \sum_{i} \frac{\int_{\Delta M_i} \psi(M) dM}{\int_{8}^{100} \psi(M) dM} f_i(E)~,
\end{equation}
where $\Delta M_i$ is the mass range of mass bin $i$, $\psi(M) = dn/dM$ is the IMF, and $f_i(E)$ is the spectrum as defined in equation (\ref{eq:pinchedFD}) with neutrino spectral parameters $(E_\nu^{\rm tot}, \langle E_\nu \rangle, \langle \alpha \rangle)$ for the core collapse of a progenitor of mass $M_i$. Since we characterize the spectral parameters using the compactness of the progenitor, yet the abundance of progenitors is determined by its ZAMS mass, we require knowledge of the distribution of compactness in mass, $\xi(M)$. We adopt the distribution from the pre-supernova models of WHW02, but explore other possibilities later. Finally, we adopt a Salpeter IMF with $\psi(M) \propto M^\eta$ with $\eta = -2.35$ in the mass range $8$--$100 M_\odot$, but explore a liberal range from $-2.15$ to $-2.45$ \citep{Bastian:2010di} in our final calculations. 

\begin{figure}
\includegraphics[width=125mm,bb=0 50 800 580]{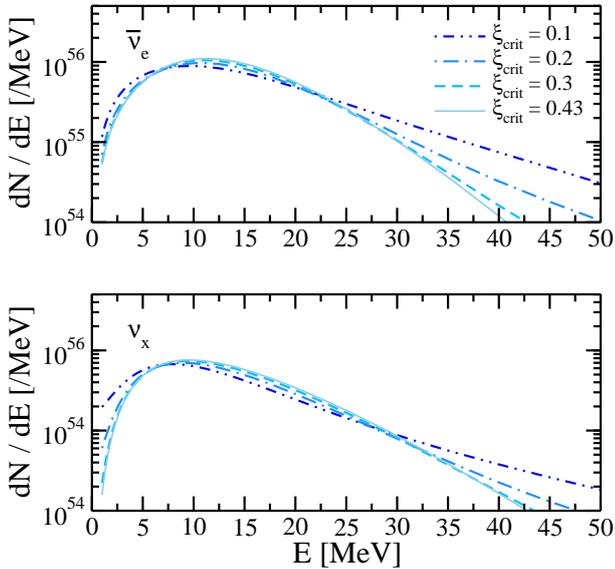}
\caption{Weighted average neutrino spectra of $\bar{\nu}_e$ (top panel) and $\nu_x$ (bottom panel), based on 101 2D core-collapse simulations and a collection of simulations of collapse to black holes. The relative contributions from neutron star and black hole scenarios are determined by the critical compactness, $\xi_{\rm 2.5,crit}$; progenitors with compactness $\xi_{2.5} > \xi_{\rm 2.5,crit}$ are assumed to collapse to black holes. For reference, the fraction of black hole collapses are 45\% ($\xi_{\rm 2.5,crit}=0.1$), 17\% ($\xi_{\rm 2.5,crit}=0.2$), 5\% ($\xi_{\rm 2.5,crit}=0.3$), and 0\% ($\xi_{\rm 2.5,crit}=0.43$). Above $\xi_{\rm 2.5,crit}=0.43$, there is no black hole contribution.}
\label{fig:spectrum}
\end{figure}

The resulting average DSNB flux is shown by the solid curves in Figure \ref{fig:spectrum}, for the $\bar{\nu}_e$ (top panel) and $\nu_x$ (bottom panel). The average $\bar{\nu}_e$ spectrum can be well modeled by the pinched Fermi-Dirac spectral function with total energetics $4.3 \times 10^{52}$ erg, mean energy $14.6$ MeV, and shape parameter $3.3$. The $\nu_x$ depends on the assumed shape parameter. In the lower panel, a shape parameter of 3.0 is assumed, but we will explore values in the range 1.0--4.0 in later sections.

The contribution from collapse to black holes is introduced by considering a critical compactness, $\xi_{\rm 2.5,crit}$, above which progenitors are assumed to collapse to black holes. While this is a simplistic picture of a complex phenomenon, our prescription is motivated by various studies showing that large compactness is conducive to black hole formation \citep[e.g.,][]{O'Connor:2010tk}. In general, the precise value of $\xi_{\rm 2.5,crit}$ will depend on the explosion mechanism and their implementation. For the neutrino mechanism, current implementations suggest values between 0.2 and 0.6 \citep{O'Connor:2010tk,Ugliano:2012kq,Horiuchi:2014ska,Pejcha:2014wda,Ertl:2015rga}. However, ongoing efforts are expected to update predictions in the future. We thus treat $\xi_{\rm 2.5,crit}$ as a parameter of interest that is predicted by simulations and to be tested by future observational data. For the neutrino emission of failed explosions, we adopt the functional form equation (\ref{eq:pinchedFD}) with neutrino spectral parameters predicted by black hole forming simulations (Figure \ref{fig:nuBH}). 

The predicted weighted average neutrino spectra including contributions from collapse to black holes are shown as non-solid lines in Figure \ref{fig:spectrum}. Four values of $\xi_{\rm 2.5,crit}=0.1$, 0.2, 0.3, and 0.43 are shown. Based on the solar metallicity progenitors of WHW02, these critical values correspond to failed explosion fractions (the number of failed explosions over the total number of massive stars in the mass range 8--$100 M_\odot$) of 45\%, 17\%, 5\%, and 0\%, respectively. Note that the largest compactness in the WHW02 solar metallicity stellar suite is $\xi_{2.5} \approx 0.43$. Therefore, values of $\xi_{\rm 2.5,crit}$ above $0.43$ means no contribution from collapse to black holes in our calculations. 

The resulting spectra are the combined effect of failed explosions having lower neutrino total energies and higher mean energies than the 2D counterparts. The smaller the critical compactness, the larger the contribution from failed explosions, and thus the more prominent is the high-energy component of the mean neutrino spectrum. Adopting a shallower (steeper) IMF increases (decreases) the emphasis on the most massive stars. For example, compared to the Salpeter IMF, a slope of $-2.15$ implies $\sim 30$\% larger representation of the most massive stars $M > 40 M_\odot$. However, the most massive stars are rare and the increase is modest in absolute terms. Even for the shallow $-2.15$ slope and largest failed fraction ($\xi_{\rm 2.5,crit}=0.1$), the weighted average neutrino spectrum is only affected at the few percent level below neutrino $\sim 30$ MeV and $\sim 15$\% above $\sim 60$ MeV. 

The core compactness of massive stars has recently been carefully investigated by \cite{Sukhbold:2013yca} using 1D stellar evolution codes. They show that quantitatively, the compactness of a star depends on a range of inputs, including not only the initial stellar mass and metallicity, but also the way mass loss and convection are handled in the code, as well as the nuclear microphysics implementation. However, the authors also show that qualitatively the compactness robustly follows a non-monotonic distribution in ZAMS mass, with a peak around $\sim 20 M_\odot$. This is the result of the interplay of the carbon-burning shell with the carbon-depleted core, and later, oxygen-burning shell with the oxygen-depleted core. Nevertheless, the position of the peak has an uncertainty of some $\sim 1 M_\odot$ in mass \citep{Sukhbold:2013yca}. To explore other currently-available suites of pre-supernova progenitor models, we determine the average neutrino flux employing the pre-supernova models of \cite{Woosley:2007as}. This suite of progenitors in general has similar or higher compactness compared to WHW02, reaching a peak compactness of $\xi_{2.5} \approx 0.54$ compared to $0.43$ for WHW02. Also, a second peak in compactness at $\sim 40 M_\odot$ is evident, in addition to the peak around $\sim 20 M_\odot$ that is seen in WHW02 and \cite{Sukhbold:2013yca}. These features manifest as a harder predicted average neutrino spectra, because higher compactness yields higher neutrino luminosities and mean energies (Figure \ref{fig:nu}). In Section \ref{sec:dsnb:signal}, we show how this affects the DSNB event rate prediction.

\subsection{DSNB flux prediction}\label{sec:effects}

The DSNB is determined by integrating the cosmic history of the comoving core-collapse rate, $R_{CC}(z)$, by the mean neutrino spectrum per core collapse, $dN/dE$, appropriately redshifted, over cosmic time \citep[see, e.g.,][for a recent review]{Beacom:2010kk},
\begin{equation} \label{eq:DSNB}
\frac{d\phi}{dE} = c \int R_{\rm CC}(z)
\frac{dN}{dE^\prime} (1+z) \left| \frac{dt}{dz} \right| dz~,
\end{equation}
where $E^\prime = E (1+z)$ and $|dz/dt| = H_0 (1+z) [\Omega_m(1+z)^3 + \Omega_\Lambda]^{1/2}$. We include contributions out to a redshift of 5, which is sufficiently large to include the majority of the DSNB flux \citep{Ando:2004hc} for $\sim 10$ MeV neutrino energy threshold. 

For the cosmic history of the core-collapse rate, we adopt those predicted from the comoving star formation rate \citep[e.g.,][]{Hopkins:2006bw}, which yields a robust estimate for the core-collapse rate regardless of whether the collapse generates a luminous supernova or not \citep{Horiuchi:2011zz}. The scaling from the star formation rate to the core-collapse rate is, 
\begin{equation}\label{eq:CCrate}
R_{\rm CC}(z) = \dot{\rho}_*(z)\frac{\int_{8}^{100}\psi(M)dM}{\int_{0.1}^{100} M \psi(M)dM}~,
\end{equation}
where $\rho_*(z)$ is the cosmic star formation rate in units $M_\odot {\rm \, yr^{-1} \, Mpc^{-3} }$, which is estimated by various priors \cite[e.g., far-infrared, ultra-violet, emission lines, and others, see, e.g.,][]{Kennicutt:1998zb} as, 
\begin{equation}\label{eq:SFR}
\dot{\rho}_*(z) = f_x L_x(z)~,
\end{equation}
where $L_x$ is the observed luminosity density of the priors and $f_x$ is the conversion factor to the star formation rate. 

The DSNB prediction depends weakly on the IMF shape. The IMF is one of the most important inputs determining the values of $f_x$, with variations of close to a factor $\sim 2$ \citep{Kennicutt:1998zb,Hopkins:2004ma,Horiuchi:2013bc}. However, this is nearly fully cancelled by the ratio of integrals in equation (\ref{eq:CCrate}). This is because the massive stars used as proxies for star formation are close in mass range to core-collapse progenitors. As a result, the product changes only at the level of a few percent. 

\subsection{Neutrino mixing and DSNB detection}\label{sec:dsnb:signal}

The neutrinos that are emitted from the neutrinospheres undergo oscillations during their propagation to a terrestrial detector. The oscillation we implement is that of matter-induced MSW during propagation through the progenitor. This results in a $\bar{\nu}_e$ survival probability of $\cos^2 \theta_{12}$ for the normal mass hierarchy (NH), where $\theta_{12}$ is the solar mixing angle and $\sin^2 \theta_{12} \simeq 0.3$, and a survival probability of $\approx 0$ for inverted mass hierarchy \citep[IH;][]{Dighe:1999bi}. The terrestrial flux of $\bar{\nu}_e$ is therefore given by,
\begin{eqnarray}\label{eq:MSW}
{\rm(NH)} \quad F^{\rm obs}_{\bar{\nu}_e} &\simeq& \cos^2 \theta_{12} F_{\bar{\nu}_e} + \sin^2 \theta_{12} F_{{\nu}_x}~, \\
{\rm(IH)} \,\, \quad F^{\rm obs}_{\bar{\nu}_e} &\simeq& F_{{\nu}_x}~.
\end{eqnarray}

Additional flavour mixing can be induced by the coherent neutrino-neutrino forward scattering potential. Although a complete picture of self-induced flavour conversions under multi-angle treatment is still missing \citep[for a review, see, e.g.,][]{Duan:2010bg,Mirizzi:2015eza}, the most uncertain epoch is the accretion phase \citep[see, e.g.,][]{Chakraborty:2011nf}, which provides of the order of tens of per cent of the total neutrino flux. Self-induced effects are not important during the earlier neutronization burst because of the large excess of $\nu_e$ due to core deleptonization \citep{Hannestad:2006nj}, and are also less relevant during the later cooling phase when the different neutrino flavours tend towards similar spectra. \cite{Lunardini:2012ne} investigated the effect of self-induced flavour conversions on the time-integrated neutrino emission and found that indeed it is subdominant, affecting at some $O$(10)\% compared to the MSW effect \citep[see also, e.g.,][]{Chakraborty:2008zp}. We therefore consider only MSW effects, but bear in mind that additional oscillation effects may occur at a subdominant level. 

For detection we consider the Super-Kamiokande (SK) and Hyper-Kamiokande (HK) water Cherenkov detectors, with 22.5 kton and 374 kton inner volumes, respectively. The main backgrounds above $\sim 10$ MeV energies are associated with atmospheric neutrinos: (1) atmospheric $\bar{\nu}_e$, (2) charge-current scattering of atmospheric $\nu_\mu$ and $\bar{\nu}_\mu$ that produce sub-Cherenkov muons (so-called ``invisible muons''), (3) atmospheric neutral current scattering, and (4) neutral current inelastic scattering with pion generation \citep[see, e.g.,][and references therein]{Bays:2011si}. SK is currently undergoing preparations to upgrade its tank with gadolinium salt, which would improve signal/background differentiation by a delayed neutron-tagging of inverse-$\beta$ events \citep{Beacom:2003nk}. This would be particularly effective in reducing invisible muons, which remain the dominant background for DSNB searches \citep{Bays:2011si}. It is not determined whether the technique will be applied in HK, or whether tagging using captures on protons will be improved. We therefore conservatively adopt lepton detection energy ranges of 10--26 MeV and 18--26 MeV for SK and HK, respectively. The cross-section for the inverse-$\beta$ decay interaction in water is accurately known \citep{Vogel:1999zy,Strumia:2003zx}. Other large-volume detectors such as JUNO, DUNE, and other proposals offer opportunities for complementary information \citep{Cocco:2004ac,Mollenberg:2014pwa,Wei:2016vjd}, but the expected event rates are lower than in HK and we do not consider them in this work. 

\begin{figure}
\includegraphics[width=125mm,bb=0 50 800 580]{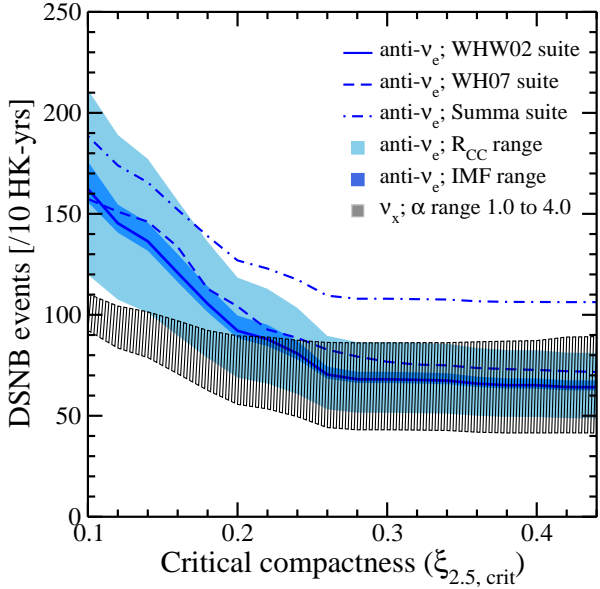}
\caption{Predicted DSNB event rate per 10 yr in HK (374 kton inner volume) as functions of the critical compactness $\xi_{\rm 2.5, crit}$, for full $\bar{\nu}_e$ survival (blue) and no $\bar{\nu}_e$ survival (black). In general, oscillations will mix these two, the exact ratios depending on scenario (e.g., the MSW mixing is shown in Figure \ref{fig:dsnb:msw}). Shown are the variations of our calculations due to (i) the core-collapse rate (light blue band), (ii) the IMF (dark blue band), (iii) pre-supernova progenitor compactness (blue dashed line), (iv) core-collapse simulation setup (blue dot--dashed line) and (v) the spectral shape parameter of $\nu_x$ (black band). The uncertainties shown for full $\bar{\nu}_e$ survival (blue bands and curves) will equally apply for no $\bar{\nu}_e$ survival (black), but are not shown for visual clarity.}
\label{fig:dsnb}
\end{figure}

\begin{figure}
\includegraphics[width=125mm,bb=0 50 800 580]{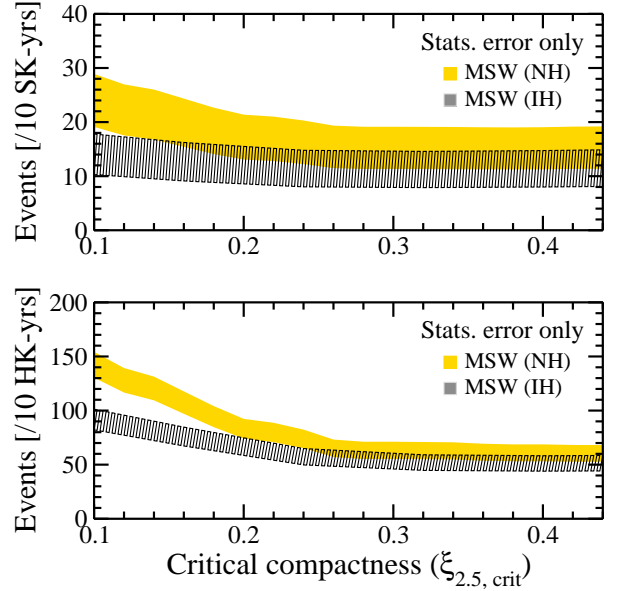}
\caption{The same as Figure \ref{fig:dsnb}, but showing the predictions for MSW mixing implementation for 10 yr in SK (22.5 kton inner volume, top panel) and 10 yr in HK (374 kton inner volume, bottom panel), with each bar width reflecting the statistical square-root ${N}$ error only, adopting the predictions using the WHW02 suite of progenitors and assuming $\nu_x$ shape parameter of 3.0. In all predictions, small critical compactness leads to more massive stars collapsing to black holes, thereby increasing the DSNB event rate. Apart from a simple normalization, slight differences appear between the SK and HK dependences on $\xi_{\rm 2.5, crit}$, due to the different detection threshold.}
\label{fig:dsnb:msw}
\end{figure}

In Figures \ref{fig:dsnb} and \ref{fig:dsnb:msw} we show the total predicted DSNB event rates as functions of the critical compactness, $\xi_{\rm 2.5,crit}$. As expected, decreasing $\xi_{\rm 2.5,crit}$ increases the predicted DSNB event rate because of the larger contribution from failed explosions especially in the higher energy range (see Figure \ref{fig:spectrum}). 

In Figure \ref{fig:dsnb} we show some of the important model prediction uncertainties. For this purpose we consider full $\bar{\nu}_e$ survival (shown in blue) and no survival (i.e., $ \bar{\nu}_e^{\rm obs} = \nu_x$, shown in black). We assume the HK detector. The uncertainty bands due to uncertain cosmic core-collapse rate and the uncertain $\nu_x$ shape parameter between 1.0 and 4.0 are large, but both are expected to be dramatically reduced with more observational and theoretical studies in the near future. We also show the effects of varying the IMF slope between $-2.15$ and $-2.45$. The IMF affects the DSNB via different weights to progenitors, as well as through different core-collapse rates, but the combined effects on the DSNB are small. As a separate prediction curve (not a band) we also show the results of adopting the pre-supernova progenitor models of \cite{Woosley:2007as}. The \cite{Woosley:2007as} progenitors have higher compactness compared to WHW02, which results in $\sim 10$\% larger DSNB event rates. Finally, we show the results based on the 18 core-collapse models of \cite{Summa:2015nyk}, which are $\sim 60$\% larger than those based on \cite{Nakamura:2014caa}. The simulations of \cite{Summa:2015nyk} typically predict larger $\bar{\nu}_e$ total energetics, higher mean energies, and smaller shape parameters (see Figure \ref{fig:nu2}). Although the differences are small, their combined effects result in a noticeable effect for the DSNB event rates, which are biased towards the high-energy portion of the neutrino emission. The large difference highlights future needs of improved simulation suites extending to late times. However, we should caution that we have taken more assumptions in our treatment of the \cite{Summa:2015nyk} models, namely, in the extrapolation to late-times. For the present purpose, we stress the fact that the predicted DSNB event rates show a similar rise with smaller critical compactness, turning around $\xi_{2.5,{\rm crit}} \sim 0.25$.

The panels in Figure \ref{fig:dsnb:msw} show results after MSW oscillations, for SK (top panel) and HK (bottom panel). The error bands show root-$N$ errors only. The predicted rates rise with smaller $\xi_{\rm 2.5,crit}$, but the uncertainty in SK is large, even after 10 yr operation. The predictions for SK and HK show some subtle shape differences, which arises due to the different detection thresholds. Since HK is proportionally more dependent on the higher energy neutrinos, the higher mean energies of $\nu_x$ become relatively more important. HK, with sufficient event statistics, can test small values of the critical compactness. 

\subsection{Impact of nuclear EOS}\label{sec:uncertainties:eos}

In addition to the structure of the progenitor, the EOS of dense matter strongly impacts predictions of the neutrino signals in the collapse to black holes. While the compactness in large part is reflective of the accretion rate and thus the evolution of the mass of the protoneutron star, the EOS determines the critical mass of the fattening central compact object at which dynamical collapse occurs. The maximum mass that can be supported depends on the stiffness of the EOS and the amount of trapped leptons, as well as the temperature in the accreting protoneutron star. A soft EOS leads to a compact protoneutron star with high density and temperature. This may lead to a high accretion luminosity of $\nu_e$ and $\bar{\nu}_e$ from the gravitational energy release and a high mean energy of $\nu_\mu$ and $\bar{\nu}_{\mu}$ due to the diffusion from the high temperature core.  

For example, the critical mass of the protoneutron star ranges between a lower value of $2.1M_{\odot}$ in baryon mass for the case of LS EOS and up to $2.7M_{\odot}$ for the stiffer Shen EOS. This uncertainty results in a factor of $\sim 2$ difference in the time duration till the black hole formation \citep{sum06,sum07,O'Connor:2010tk}, and therefore, different values for the total neutrino energies of the black hole component in the prediction of the DSNB. Indeed, adopting the stiffer Shen EOS, \cite{Lunardini:2009ya} and \cite{Keehn:2010pn} have shown that the failed explosion contribution can double the predicted DSNB events, when a failed fraction of $0.22$ is adopted. In fact, as long as other uncertainties can be controlled, this would open the intriguing possibility to assess the EOS from the DSNB. 

Further investigations of the DSNB by focusing on the influence of EOS are clearly needed. In particular, it is important to consider the outcome using new sets of EOS tables \citep[e.g.,][]{2016arXiv161003361O}. These EOSs are carefully tuned to match symmetry energy measurements by recent nuclear experiments, neutron star radii extracted from astronomical observations, the maximum mass of cold neutron stars from binary observations, and other data. These constraints are marginally fulfilled by LS 220 as well. 

In order to investigate potential effects due to other EOSs satisfying current data and constraints, we simulate the core collapse of the $40 M_\odot$ progenitor of WW95 ($\xi_{2.5} = 0.59$) with the SFHo EOS \citep{Steiner:2012rk}. We find that the time-integrated neutrino spectra are very similar to those predicted by the LS 220 assumed throughout this work. For example, the total neutrino energies are $6.0 \times 10^{52} \, {\rm erg}$, $5.1 \times 10^{52} \, {\rm erg}$, and $2.5 \times 10^{52} \, {\rm erg}$ for $\nu_e$, $\bar{\nu}_e$, and $\nu_x$, respectively. By comparison, the equivalent in LS220 are $6.3 \times 10^{52} \, {\rm erg}$, $4.9 \times 10^{52} \, {\rm erg}$, and $2.4 \times 10^{52} \, {\rm erg}$. The differences between the two EOS are smaller than the variations due to the range of compactness considered in this paper (Figure \ref{fig:nuBH}). The situation is mirrored in the mean energies and shape parameter: in SFHo they are 13.4 MeV, 16.1 MeV, and 18.2 MeV and 3.0, 3.3, and 1.8, for $\nu_e$, $\bar{\nu}_e$, and $\nu_x$, respectively, close to LS220 values (Figure \ref{fig:nuBH}). While these initial results are encouraging, it will be important in the future to explore the numerical impacts of a larger range of EOS tables.

\section{Discussions and conclusions}\label{sec:discussion}

We have revisited predictions of the diffuse supernova neutrino background (DSNB) in light of recent suites of core-collapse simulations starting with large numbers of progenitor initial conditions. We have separately taken into account the contribution from three channels: core collapse of ONeMg cores, Fe core collapse to neutron stars, and core collapse to black holes. In particular, for the latter two channels we accommodated the variations in the neutrino emission due to progenitor structure by using the compactness parameter determined at the mass coordinate of $2.5 M_\odot$ (see equation \ref{eq:compactness}) and over 100 core-collapse simulations. To this end, we first characterized the neutrino spectral parameters total energy $E_\nu^{\rm tot}$, mean energy $\langle E_\nu \rangle$, and shape parameter $\langle \alpha \rangle$ in terms of the progenitor compactness, and found noticeable dependencies (Figure \ref{fig:nu}). We then incorporated these variations in our IMF weighted DSNB flux predictions (Figure \ref{fig:spectrum} solid curves). 

We included contributions from core collapse to black holes by progressively replacing the core collapse to neutron stars with collapse to black holes, starting with progenitors with the largest compactness. In other words, we assume that massive stars with compactness above some threshold value all collapse to black holes, and stars with smaller compactness all collapse to neutron stars. Different values of the threshold critical compactness therefore lead to different numbers (non-monotonic in progenitor mass) of massive star explosions being replaced by collapse to black holes and affect the DSNB flux (Figure \ref{fig:spectrum} non-solid curves). 

Our choice to use the compactness as a suitable variable is based on its utility to characterize the dependence of the mass accretion on the progenitor. We have opted for a $2.5 M_\odot$ for our definition of compactness, which works particularly well for progenitors with a relatively delayed explosion. Our choice is also motivated by its utility as a proxy for whether a progenitor collapses to a black hole or not \citep{O'Connor:2010tk}. However, this is a simplified treatment with limitations. Studies have shown that no single critical compactness is 100\% accurate in predicting the outcome \citep{Ugliano:2012kq,Sukhbold:2013yca,Sukhbold:2015wba,Ertl:2015rga}. For example, progenitors with faster shock revivals are better characterized by compactness defined by a smaller mass, e.g., $1.5 M_\odot$. Nevertheless, a single definition of compactness is able to predict the outcome in 80--90\% of cases \citep[e.g.,][]{Pejcha:2014wda}. From a theoretical perspective, the critical compactness depends on which progenitor condition leads to explosions, i.e., it depends on the explosion mechanism and its implementation. The neutrino-driven delayed explosion mechanism applied across multiple progenitors suggest values in the range $\xi_{2.5,{\rm crit}}=0.2$--0.6  \citep{O'Connor:2010tk,Ugliano:2012kq,Horiuchi:2014ska,Pejcha:2014wda,Ertl:2015rga}. The large range is partly due to limitations in current suites of core-collapse simulations, i.e., 1D simulations require artificially increased neutrino heating, 2D simulations are artificially conducive to explosions, and 3D simulations are computationally expensive. Future simulations will address these concerns and improve predictions, making it interesting to explore their testability. 

We showed that even the absolute event rate of DSNB in future neutrino detectors will have the power to discriminate between values of the critical compactness realized in nature (Figure \ref{fig:dsnb:msw}), once major uncertainties are reduced (Figure \ref{fig:dsnb}). A spectral fit analysis will be possible in the era of high-statistics DSNB events, and should yield even better information. Sensitivity is particularly strong in the region of small critical compactness. Such small values imply a high fraction (larger than $\sim 20$\%) of massive stars undergoing collapse to black holes. Interestingly, such large fractions remain a possibility by current observations. For example, searches for the disappearance of massive stars \citep{Kochanek:2008mp} have identified one candidate collapse to a black hole \citep{Gerke:2014ooa}, which leads to a fraction of failed explosions of 4--43\% at 90\% C.L.~\citep{Adams:2016hit}. Furthermore, pre- and post- observations of the black hole candidate reveal an object consistent \citep{Adams:2016ffj} with predictions of collapse to a black hole \citep{Lovegrove:2013hox}. Other observational discrepancies, from the ratio of massive star formation and explosion \citep{Horiuchi:2011zz} to the red supergiant problem \citep{Smartt:2008zd,Smartt:2015sfa}, can also be explained with a critical compactness of $\xi_{2.5,{\rm crit}} \sim 0.2$ \citep{Horiuchi:2014ska}; and the connection to the DSNB has been explored by \cite{Hidaka:2016zei}. A similarly low critical compactness can also explain the mass function of compact objects \citep{Kochanek:2013yca,Kochanek:2014mwa}. It is therefore of interest to observationally test the region of small critical compactness with DSNB neutrinos.

We showed that among the largest uncertainty is the historical core-collapse rate, but measurements are improving rapidly \cite[e.g., for a sense of the improvement in the data in the several years, compare the compilations in][]{Strigari:2005hu,Horiuchi:2008jz} and further improvements are expected by upcoming wide-field surveys such as LSST \citep{Lien:2010yb}. Another large uncertainty comes from the details of the simulation setup. Using the suite of 2D simulations of \cite{Summa:2015nyk}, we find a $\sim 60$\% increase in the predicted DSNB rate. It is clear that DSNB predictions will need to be regularly updated as core-collapse simulations improve. Nevertheless, despite the updates, the qualitative trend of the DSNB with the critical compactness will likely hold. The DSNB event predictions also depend on modeling of stellar evolution, i.e., the distribution of compactness in massive stars. Ongoing investigations of stellar evolution including improved treatments of semiconvection and multi-dimensional effects will be important to shed light on this uncertainty \citep[e.g., see discussions in][]{Sukhbold:2013yca}. The final strong dependence is on the neutrino mass hierarchy (Figure \ref{fig:dsnb:msw}), but we expect this can be tested by alternate means, e.g., long baseline experiments \citep[e.g.,][]{Abe:2011ts,Adams:2013qkq} and/or a future Galactic core-collapse neutrino burst detection \citep[e.g.,][]{Mirizzi:2015eza}. Throughout, we have neglected exotic neutrino properties such as neutrino magnetic moment, neutrino decays, or extra sterile neutrino states, which may impact the neutrino signal depending on scenarios and parameters \citep[e.g.,][]{Barbieri:2000mg,Hidaka:2006sg,Hidaka:2007se,Heger:2008er,Fuller:2009zz,Baerwald:2012kc,deGouvea:2012hg}. Some of these will be constrained by future neutrino experiments and future Galactic core-collapse neutrino detections. 

The DSNB is a guaranteed flux of core-collapse neutrinos that is close to detector sensitivity. Planned detector upgrades and new analysis techniques will improve background rejection \cite[e.g.,][]{Beacom:2003nk,Li:2015kpa,Li:2015lxa} and are anticipated to help deliver the first detection of the DSNB. Future megaton scale neutrino detectors will usher in an era of high-statistics DSNB detections that allow for detailed probes of the DSNB inputs. We have explored theoretically the variations in the neutrino emission due to progenitor structure, including the transition from core collapse that leave behind neutron stars to those that leave behind black holes. As core-collapse simulations improve and the core-collapse rate become better known, DSNB predictions will continue to improve and allow future DSNB detections to reveal the diversity in massive stellar core collapse. 

\section*{Acknowledgments}

We thank Ken'ichiro Nakazato for helpful discussions and sharing progenitor information of their simulations. We thank the referee, Raphael Hirschi, for very carefully reading the manuscript and providing helpful comments. SH is supported by the US Department of Energy under award number de-sc0018327. We acknowledge the usage of computing resources at KEK, JLDG, RCNP, YITP and UT. This study was supported in part by the Grants-in-Aid for the Scientific Research of Japan Society for the Promotion of Science (JSPS, Nos.~24244036, 26707013, 26870823, 15K05093, 16K17668, 17H06364), the Ministry of Education, Science and Culture of Japan (MEXT, Nos.~26104001, 26104006, 15H00789, 15H01039, 15KK0173, 17H01130, 17H05205, 17H06357, 17H06365), and JICFuS as a priority issue to be tackled by using Post `K' Computer. KN and KK are supported by the Central Research Institute of Fukuoka University (Nos.171042, 177103). TF acknowledges support from the Polish National Science Center (NCN) under grant number UMO-2016/23/B/ST2/00720. This article is based upon work from the ``ChETE'' COST Action (CA16117), supported by COST (European Cooperation in Science and Technology). At Garching, this work was supported by the European Research Council through grant ERC AdG 341157-COCO2CASA and by the Deutsche Forschungsgemeinschaft through grants SFB 1258 (``Neutrinos, Dark Matter, Messengers'') and EXC 153 (``Excellence Cluster Universe''). The data of the Garching simulations are available for download upon request at {\tt http://wwwmpa.mpa-garching.mpg.de/ccsnarchive/archive.html}. 




\bibliographystyle{mnras}
\bibliography{ms} 








\bsp	
\label{lastpage}
\end{document}